\documentclass[12pt,twoside,fleqn]{article}

\usepackage{epsfig}
\usepackage{psfig}

\def\lsim{\raise0.3ex\hbox{$<$\kern-0.75em\raise-1.1ex\hbox{$\sim$}}}
\def\gsim{\raise0.3ex\hbox{$>$\kern-0.75em\raise-1.1ex\hbox{$\sim$}}}
\makeatletter
\@addtoreset{equation}{section}
\makeatother

\newcommand{\beqn} {\begin{equation}}
\newcommand{\eqn} {\end{equation}}

\newcommand{\slsh}[1] {#1\kern-.43em/}

\newcommand{\real}{{\sf I}\kern-.12em{\sf R}}
\newcommand{\comp}{{\sf I}\kern-.48em{\sf C}}
\newcommand{\nin} {\in\kern-.6em/}

\def\ie{{i.e.\/}}

\def\MEF{m_{\rm eff}}\def\mef{\ifmmode\MEF\else$\MEF$\fi}

\def\NT{N_\tau}
\def\nt{\ifmmode\NT\else$\NT$\fi}
\def\NS{N_\sigma}
\def\ns{\ifmmode\NS\else$\NS$\fi}

\def\PRL{{ Phys.\ Rev.\ Lett.\ }}
\def\PL{{ Phys.\ Lett.\ }}
\def\NP{{ Nucl.\ Phys.\ }}

\def\MO{{\langle |L| \rangle }}

\begin{document}
\thispagestyle{empty}
%
%\hbox{}
 \mbox{} \hfill BI-TP 96/04~~\\
 \mbox{} \hfill January 1996\\
\begin{center}
\vspace*{1.0cm}
{{\large \bf Thermodynamics of SU(3) Lattice Gauge Theory}
 } \\
\vspace*{1.0cm}
{\large G. Boyd, J. Engels, F. Karsch, E. Laermann, C. Legeland, \\
M. L\"utgemeier and B. Petersson} \\
\vspace*{1.0cm}
{\normalsize
$\mbox{}$ {Fakult\"at f\"ur Physik, Universit\"at Bielefeld,
D-33615 Bielefeld, Germany}}\\
\vspace*{2cm}
{\large \bf Abstract}
\end{center}
\setlength{\baselineskip}{1.3\baselineskip}

The pressure and the energy density of the $SU(3)$ gauge theory are
calculated on lattices with temporal extent $N_\tau = 4$, 6 and 8 and spatial
extent $N_\sigma =16$ and 32. The results are then extrapolated to
the continuum limit. In the investigated temperature range up to five times
$T_c$ we observe a $15\%$ deviation from the ideal gas limit.
We also present new results for the critical temperature on
lattices with temporal extent $N_\tau = 8$ and 12.
At the corresponding critical couplings the string tension is calculated
on $32^4$ lattices to fix the temperature scale. An extrapolation to
the continuum limit yields $T_c/\sqrt{\sigma} = 0.629(3)$.
We furthermore present results on the electric and magnetic condensates
as well as the temperature dependence of the spatial string tension. These
observables suggest that the temperature dependent running coupling remains
large even at $T\simeq 5T_c$. For the spatial string tension we find
$\sqrt{\sigma_s}/T = 0.566(13) g^2(T)$ with $g^2(5T_c) \simeq 1.5$.

\newpage
\setcounter{page}{1}

%%%%%%%%%%%%%%%%%%%%%%%%%%%%%%%%%%%%%%%%%%%%%%%%%%%%%%%%%%%%%%%%%%%%%%%%%%%%%

\section{Introduction}

%%%%%%%%%%%%%%%%%%%%%%%%%%%%%%%%%%%%%%%%%%%%%%%%%%%%%%%%%%%%%%%%%%%%%%%%%%%%%

The calculation of the equation of state of QCD is one of the central goals
of lattice simulations at finite temperature. It has been pursued ever since
the first finite temperature Monte Carlo calculations \cite{first}. Besides
knowing details about the QCD phase transition like the order and the
critical temperature it is of great importance for any investigation of the
QCD plasma phase to get a quantitative handle on the equation of state at
high temperatures. The intuitive picture of the high
temperature phase as a gas of weakly interacting quarks and gluons is based
on leading order perturbation theory. However,
the well-known infrared problems of perturbative QCD \cite{Linde}
lead to a poor convergence of the corresponding expansion of
the thermodynamic potential even at rather high temperatures
\cite{Arnold,Zhai}. Although the newly developed
techniques of resummed perturbation theory have led
to much progress in perturbative calculations \cite{Braaten}, it seems that
non-perturbative effects still dominate the equation of state in the
temperature regime attainable in future heavy ion experiments.

Lattice calculations of energy density, pressure and other thermodynamic
variables have already led to some insight into the temperature dependence
of these quantities in the QCD plasma phase. The energy density
in the pure gauge theory, for instance,
has been found to rise rapidly at $T_c$ and approach the high
temperature ideal gas limit from below.
However, except for a very recent
calculation for the $SU(2)$ gauge theory \cite{Eng95}, all studies of
the QCD equation of state (EOS)
have been performed on rather small lattices. The crucial limitation comes
here from the extent of the lattice in Euclidean time ($N_\tau$). So far all
calculations of bulk thermodynamic quantities have been restricted to
lattices with only four sites in the Euclidean time direction ($N_\tau =
4$) \cite{earlier}. The small extent of the lattice in the
time direction causes large lattice artifacts in thermodynamic quantities.
This comes about because the discretization of the field strength tensor
in the standard Wilson formulation of lattice QCD \cite{Wil74}
introduces $O(a^2)$ deviations from its continuum counterpart,
\ie~$O((aT)^2\equiv N_\tau^{-2})$ corrections at a fixed finite temperature $T$.
In the case of an ideal bose gas -- the perturbative high temperature
limit of an $SU(N)$ gauge theory -- the deviations of bulk thermodynamic
quantities like energy density and pressure from the corresponding
continuum expressions are as large as 50\% on lattices with only
four sites in the time direction. These effects are even
larger for a fermi gas \cite{Eng82}.

In order to compare lattice calculations of the EOS with
continuum perturbation theory or phenomenological models like the bag EOS,
it is mandatory to remove the lattice artifacts.
This requires a systematic analysis of
thermodynamic quantities on lattices with varying $N_\tau$.
At fixed temperature $T$, one can then extrapolate the numerical results
to the continuum limit $a \sim 1/N_\tau \rightarrow 0$.
The methods required to achieve this have been
developed over last few years. In particular, it is now possible to
calculate the energy density and pressure
non-perturbatively, without using certain perturbative
approximations as in earlier calculations \cite{Eng90}. However,
since these quantities are measured on the lattice in units of the
lattice spacing, e.g. $\epsilon a^4 \sim N_{\tau}^{-4}$, the statistics
required for constant accuracy increases with $N_\tau$ correspondingly,
\ie~ $N_{\tau}$ cannot be chosen too large.

In this paper we present a systematic study of the thermodynamics of the
pure gauge sector of QCD, \ie~we neglect the contribution of fermions
and study the thermodynamics of a gluon gas, which is described by an
$SU(3)$ gauge theory. We calculate thermodynamic quantities on
lattices of size $N_\sigma^3 \times N_\tau$ with $N_\tau = 4$, 6 and 8.
In addition, we perform calculations of the string tension on large
symmetric lattices of size $32^3 \times 32$,
\ie~at zero temperature, in order to
set a physical scale for the temperature in our calculations. The results
for different values of $N_{\tau}$ are then used to extrapolate to the continuum
limit. In the next section we will outline the basic formalism for our
calculations. In section 3 we discuss our results for the critical couplings
of the deconfinement transition and the string tension at zero temperature.
These are used in section 4 to determine the temperature scale for the
thermodynamic observables.
In section 5 we present our results for pressure, energy density and other
thermodynamic quantities and describe their extrapolation to the continuum
limit. A discussion of the results and a comparison with continuum models is
given in section 6. Part of these results have already been reported in Ref.~
\cite{eosprl}. Furthermore in section 6 results for the magnetic and 
electric condensates as well as for the spatial string tension are presented.

%%%%%%%%%%%%%%%%%%%%%%%%%%%%%%%%%%%%%%%%%%%%%%%%%%%%%%%%%%%%%%%%%%%%%%%%%%%%

\section{Thermodynamics on the Lattice}

%%%%%%%%%%%%%%%%%%%%%%%%%%%%%%%%%%%%%%%%%%%%%%%%%%%%%%%%%%%%%%%%%%%%%%%%%%%%

To start the discussion of the thermodynamics of $SU(N)$ gauge theories
on the lattice we recall some basic thermodynamic relations in the
continuum. All thermodynamic quantities can be derived from the
partition function $Z(T,V)$. Its logarithm defines the free energy density,
\beqn
f  = - {T \over V} {\rm ln} Z(T,V)~~.
\label{freeenergy}
\eqn
The energy density and pressure
are derivatives of $\ln Z$ with respect to $T$ and
$V$,
\begin{eqnarray}
\epsilon&=&{T^2 \over V} {\partial {\rm ln} Z(T,V)
\over \partial T}~~,~~\label{energy} \\%\nonumber \\
p&=&T {\partial {\rm ln} Z(T,V) \over \partial V}~~.~~
\label{druck}
\end{eqnarray}
For large, homogeneous systems, however, the pressure can directly be
obtained from the free energy density,
\beqn
p = -f~~.   
\label{pressure}
\eqn
Using this relation one can express the entropy density
$s=(\epsilon+p)/T$ and the difference between $\epsilon$ and $3p$
in terms of derivatives of the pressure with respect to temperature,
\begin{eqnarray}
{\epsilon + p \over T} & = & {\partial p \over \partial T } ~,~~
\label{entropy} \\
{\epsilon - 3p } & = & T^5 {\partial \over \partial T} (p/T^4) ~~.
\label{delta}
\end{eqnarray}

On the lattice, temperature and volume of the thermodynamic
system are determined by the lattice size, $N_\sigma^3 \times N_\tau$,
and the lattice cut-off $a$,
\beqn
T^{-1} = N_\tau a \qquad , \qquad V = (N_\sigma a)^3 ~~.
\label{TandV}
\eqn
In an $SU(N)$ gauge theory the lattice cut-off $a$ is a function of the
bare gauge coupling $\beta \equiv 2N/g^2$. This function
fixes the temperature and the physical volume at a given coupling and
is known to two-loop order in perturbation theory.
To obtain it non-perturbatively one calculates a physical observable
with non--trivial dimension at various values of $\beta$ on the lattice,
\ie~in units of the cut-off.
In pure $SU(3)$ gauge theory
the string tension $\sigma a^2$ or the critical temperature $T_c a$ for
the deconfinement transition are suitable observables. This
then defines the temperature scales $T/\sqrt{\sigma}$ or $T/T_c$.

Although in principle all thermodynamic quantities can be derived from
the free energy density, in practice a direct
computation of the partition function on the lattice is
not possible. A way out is to calculate the expectation value
of the action, \ie~the derivative of $\ln Z$ with respect to
the bare gauge coupling $\beta$. Up to a normalization constant, resulting
from the lower integration limit $\beta_0$, the free
energy density is then obtained by integrating this expectation value
\beqn
{f\over T^4}\Big\vert_{\beta_0}^{\beta} =~-N_\tau^4\int_{\beta_0}^{\beta}
{\rm d}\beta' \bigl[ S_0-S_T \bigr]~,
\label{freelat}
\eqn
with $S_0=6 P_0$ and $S_T= 3(P_\tau + P_\sigma)$. Here
$P_{\sigma,\tau}$ denote the expectation values of space-space and
space-time plaquettes,
\beqn
P_{\sigma,\tau} = 1-
{1 \over N} {\rm Re} \langle {\rm Tr} (U_1U_2U_3^{\dagger}U_4^{\dagger})
\rangle~,
\label{plaqui}
\eqn
on asymmetric lattices of size $N_\sigma^3 \times N_\tau$.
Similarly, $P_0$ is the plaquette expectation value
on symmetric lattices of size $N_\sigma^4$.
It serves to normalize the free energy density to be zero at zero temperature.
As the physical excitations in the low temperature phase of
an $SU(N)$ gauge theory are glueballs, which are known to be quite heavy
($m_G \gsim 1$~GeV), the free energy density is expected to drop
exponentially ($f \sim \exp(-m_G/T)$). We thus assume that $f$ is negligible
below a temperature quite close to $T_c$. The
lower integration limit $\beta_0$ can then be chosen correspondingly.

In order to relate
the free energy density to the pressure via Eq.~\ref{pressure},
one has to insure that the spatial
volume is large enough . Earlier studies for
the $SU(2)$ gauge theory \cite{Eng90} suggest that lattices with spatial
size $TV^{1/3}\equiv (N_\sigma / N_\tau) \ge 4$ are sufficient for this.
The difference between $\epsilon$ and $3p$ can be derived
from Eqs.~\ref{pressure}, \ref{delta} and \ref{freelat} as
\beqn
{\epsilon - 3p \over T^4} = N_\tau^4 T{d\beta \over dT}
[ S_0-S_T ]~~.
\label{deltalat}
\eqn
The energy density ($\epsilon/T^4$) is then obtained by adding the result
for $3p/T^4$.
On isotropic lattices with fixed temporal extent $N_\tau$ the
variation of the temperature is directly related to the variation of the
lattice spacing (Eq.~\ref{TandV}). The derivative of the bare coupling with
respect to $T$, occurring in Eq.~\ref{deltalat}, thus is connected to the
renormalization group equation for the bare coupling,
\beqn
\tilde{\beta}(g) \equiv T{d\beta \over dT} = -
a{d\beta \over da} = -2Na{dg^{-2} \over da}~~~.
\label{rgeqn}
\eqn
We will discuss this relation in the next section.

%%%%%%%%%%%%%%%%%%%%%%%%%%%%%%%%%%%%%%%%%%%%%%%%%%%%%%%%%%%%%%%%%%%%%%%%%%%%%%

\section{Critical Temperature and the Temperature Scale}

%%%%%%%%%%%%%%%%%%%%%%%%%%%%%%%%%%%%%%%%%%%%%%%%%%%%%%%%%%%%%%%%%%%%%%%%%%%%%%

\subsection{Critical Temperature and the String Tension}

%%%%%%%%%%%%%%%%%%%%%%%%%%%%%%%%%%%%%%%%%%%%%%%%%%%%%%%%%%%%%%%%%%%%%%%%%%%%%%
In order to express thermodynamic quantities, calculated on the lattice,
in physical units we have to find the relation between the
gauge coupling $\beta$ and the lattice cut-off $a$. This can, for instance,
be achieved through a calculation of the critical couplings for the
deconfinement transition $\beta_c$ on lattices with given temporal
extent $N_\tau$ or a calculation of the string tension $\sqrt{\sigma}a$ at
zero temperature. We will present here some new results for both quantities
and discuss the determination of the temperature scale from these new
calculations as well as further information from MCRG calculations
\cite{Akemi} in the next subsection.

We have calculated the critical couplings for the deconfinement phase
transition  on lattices of size $16^3 \times 4$ and $32^3 \times N_\tau$
with $N_\tau =6$, 8 and 12 .  While the critical couplings on lattices
with $N_\tau =4$ and 6 have been studied with high accuracy in the past
\cite{Kuti,Christ,Iwa92} those for the larger lattices are known only with
much less accuracy. In particular, they have been extracted from calculations
on lattices with rather small spatial extent \cite{Kuti,Christ} and are 
therefore expected to be shifted to larger values on the $N_\sigma=32$ 
lattices used here.

The critical couplings are determined from an analysis of the
Polyakov loop susceptibility,
\beqn
 \chi_L = \ns^3(\langle |L|^2 \rangle  - \MO^2)~~,
\label{suscept}
\eqn
with $L$ denoting the lattice average of the Polyakov loop,
\ie~the product of link variables
in the time direction% $\hat{0}$,
\beqn
L = {1\over N_\sigma^3} \sum_{\vec{n}} {1\over N}{\rm Tr}
\prod_{i=1}^{N_\tau} U_{(\vec{n},i),\hat{0}}~~.
\label{polyakov}
\eqn

The Polyakov loop has been determined at
four values of $\beta$ in the vicinity of the critical point. 
At each value of
the coupling we have performed between 20.000 and 30.000 sweeps -
for details of the simulation see the Appendix. Close
to the critical point one observes frequent flips between the confined and
deconfined phase. In particular for $N_\tau =8$ we find clear two-peak
structures in the Polyakov-loop distribution functions, which
support the first order nature of the transition.

\begin{figure}[htb]
\begin{center}
  \epsfig{bbllx=70,bblly=270,bburx=540,bbury=580,
       file=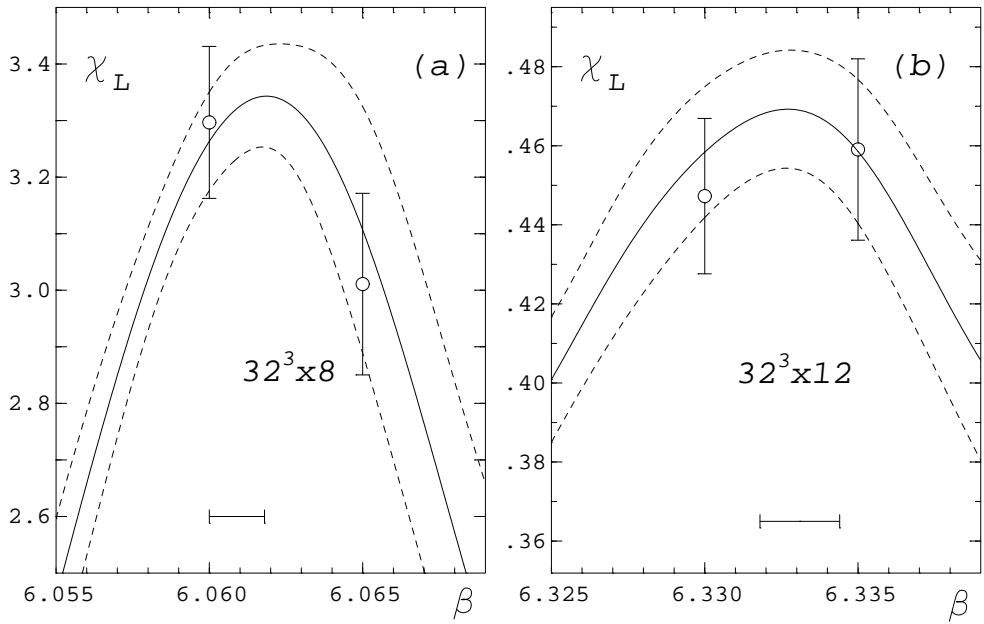, height=90mm}
\end{center}
\caption{The Polyakov-loop susceptibility on lattices of size $32^3 \times
8$ (a) and $32^3\times 12$ (b). The solid curves are the DSM interpolations,
the dashed lines show the jackknife error bands. The interpolations are
based on data collected at $\beta= 6.04$, 6.06, 6.065, 6.07 and 6.08 (a) and
$\beta= 6.32$, 6.33, 6.335 and 6.34 (b),
respectively.}
\label{fig:suscept}
\end{figure}

To determine the critical point we have interpolated the results at
the four couplings with the density of states method (DSM) \cite{Ferrenberg}.
Pronounced peaks in the Polyakov loop susceptibilities are visible
for all four lattice sizes.
In Figure~\ref{fig:suscept} we show the susceptibilities for the two larger
lattices, $32^3\times 8$ and $32^3\times 12$, in a narrow
region around the peak location.
The critical couplings and their errors have been determined from a
jackknife analysis for the maxima of the susceptibilities.
These errors are shown as horizontal bars in Figure~\ref{fig:suscept},
all critical couplings are listed in Table~\ref{tab:critical}.

\vskip 1truecm
\begin{table}
\begin{center}
\begin{tabular}{|r|c|l|l|}
\hline
$N_\tau$ & $N_\sigma$ & \multicolumn{1}{|c|}{$\beta_c ( N_\tau, N_\sigma )$ }
& \multicolumn{1}{|c|}{$\beta_c ( N_\tau, \infty)$} \\
\hline
    4    &     16     &  5.6908 ~(2) & 5.6925~(2)   \\
    6    &     32     &  5.8938 ~(11)& 5.8941~(5)   \\
    8    &     32     &  6.0609 ~(9) & 6.0625   \\
   12    &     32     &  6.3331 ~(13)& 6.3384    \\
\hline
\end{tabular}
\end{center}
\caption{Critical couplings for the $SU(3)$ deconfinement transition.
In the last column we give the extrapolations to the thermodynamic limit
($N_\sigma \rightarrow \infty$). For $N_\tau =8$ and 12 a
finite volume dependence proportional to $(N_\tau /N_\sigma )^3$ was
assumed.}
\label{tab:critical}
\end{table}

For the $N_\tau = 4$ and 6 lattices our analysis of the critical couplings is
in complete agreement with earlier determinations \cite{Iwa92}. For
$N_\tau =8$ and 12 we obtain, however, significantly larger critical
couplings than those found in previous calculations \cite{Kuti,Christ}.
The shift towards larger values is expected and
due to our larger spatial volume. In the analysis of the
critical couplings for $N_\tau=4$ and 6 a clear shift towards larger values
of $\beta_c$ has been observed with increasing spatial volume. In fact, as
the transition is first order for the $SU(3)$ gauge theory, the critical
couplings will scale like
\beqn
\beta_c (N_\tau,N_\sigma) = \beta_c (N_\tau,\infty) - h \biggl( {N_\tau \over
N_\sigma} \biggr)^3~~.
\label{bscaleing}
\eqn
Using Eq.~\ref{bscaleing} with the
constant $h\lsim 0.1$ determined from the analysis at $N_\tau = 6$
\cite{Iwa92} one can
extrapolate our critical couplings to the thermodynamic limit. This amounts
to a shift of the critical coupling by about about $0.0016$ for
$N_\tau=8$ and $0.0053$ for $N_\tau = 12$.  The resulting values
are given in the last column of Table~\ref{tab:critical}.

In order to determine an absolute scale for the temperature we have to evaluate
a physical observable which is known experimentally. A suitable quantity for
calculations in the pure gauge sector of QCD is the string tension. At the
critical couplings determined for $N_\tau = 8$ and 12 we have calculated the
string tension on lattices of size $32^4$ using standard smearing techniques
to evaluate the heavy quark potential. For this analysis we have generated
500 gauge field configurations separated by 10 sweeps. 
In these cases results for the
string tension are summarized in Table~\ref{tab:ratios} where we also give
the resulting critical temperature in units of the string tension. For
$N_\tau = 4$ and 6 the ratio $T_c/\sqrt{\sigma}$ has been evaluated at
the critical couplings extrapolated to the infinite volume limit. Results
for the string tension have been taken from Ref.~\cite{Balxx} and have been
interpolated at $\beta_c$. For
$N_\tau =8$ and 12 we evaluate this ratio at the critical couplings
determined by us on lattices with finite $N_\sigma/N_\tau$. The
volume dependence of $T_c/\sqrt{\sigma}$, resulting from a shift
of the critical couplings, is taken into account as a systematic
error, which we estimate by
assuming an exponential scaling of $\sqrt{\sigma}a$ according
to the asymptotic renormalization group equation. It is also given in
the last column of Table~\ref{tab:ratios}.

Although the ratio hardly shows a systematic cut-off dependence, we have
extrapolated the results for the different $N_\tau$-values to the
continuum limit using a fit of the form $a_0+a_2/N_\tau^2$. This yields
\beqn
{T_c \over \sqrt{\sigma}} = 0.625 \pm 0.003~(+0.004)~.
\label{Tcratio}
\eqn
The number in brackets indicates the systematic shift we expect from the
infinite volume extrapolation of the critical couplings.
Using $\sqrt{\sigma} = 420$MeV this gives a critical temperature of about
260~MeV. We note that this estimate of $T_c$ is about 10\% larger than
earlier estimates \cite{Fin93}, which is due to our newly determined critical
couplings for the larger lattices.

\begin{table}
\begin{center}
\begin{tabular}{|r|l|l|l|}
\hline
$N_\tau$  &  \multicolumn{1}{|c|}{$\beta_c$}
         &  \multicolumn{1}{|c|}{$\sqrt{\sigma} a$}
         &  \multicolumn{1}{|c|}{$T_c/\sqrt{\sigma}$}    \\
\hline
    4    &  5.6925  (2)  &  0.4179  (24)  &  0.5983  (30)       \\
    6    &  5.8941  (5)  &  0.2734  (37)  &  0.6096  (71)       \\
    8    &  6.0609  (9)  &  0.1958  (17)  &  0.6383  (55)  (+13) \\
   12    &  6.3331  (13)  &  0.1347   (6)  &  0.6187  (28)  (+42) \\
\hline
\end{tabular}
\end{center}
 \caption{The string tension at the critical couplings for the
deconfinement transition and the ratio $T_c/\protect\sqrt{\sigma}$.}
\label{tab:ratios}
\end{table}
%%%%%%%%%%%%%%%%%%%%%%%%%%%%%%%%%%%%%%%%%%%%%%%%%%%%%%%%%%%%%%%%%%%%%%%%%%%%%%

\subsection{The Temperature Scale}

%%%%%%%%%%%%%%%%%%%%%%%%%%%%%%%%%%%%%%%%%%%%%%%%%%%%%%%%%%%%%%%%%%%%%%%%%%%%%%
The calculation of a physical quantity like $\sqrt{\sigma}a$
fixes the lattice cut-off and thus
the temperature at a given value of the gauge coupling.
On a lattice with temporal extent $N_\tau$ the temperature can then be
expressed in units of the string tension and is given by
$T/\sqrt{\sigma} =1/\sqrt{\sigma}aN_\tau$. This is, in principal, sufficient
to perform all thermodynamic calculations we are going to discuss in the next
sections. In practice, however, it is much
more convenient to work with a direct parameterization of the relation
between the cut-off and the gauge coupling. This will largely simplify
the calculation of quantities that involve derivatives with respect to the
temperature (see for instance Eq.~\ref{deltalat}).

In the asymptotic scaling regime the relation between cut-off and gauge
coupling is given by the universal terms of the QCD $\beta$-function,
$a\Lambda_L = R(\beta)$, with
\beqn
R(\beta) = \biggl({\beta \over 2Nb_0} \biggr)^{b_1/2b_0^2} 
\exp [-\beta/4Nb_0]~~,
\label{renorm}
\eqn
and the coefficients, for a vanishing number of quark flavours,
\beqn
b_0={{11N}\over {48\pi^2}} \qquad , \qquad
b_1={{34}\over 3}\biggl({{N}\over {16\pi^2}}\biggr)^2~~.
\label{bees}
\eqn
However, it is well-known that in the coupling
regime close to $g^2=1$, which one is usually exploring in lattice
calculations, there are large deviations
from the asymptotic scaling relation.
Still it seems that a unique relation between the gauge coupling and the
lattice cut-off can be established in this intermediate coupling regime to
the extent that scaling violations in ratios of physical observables are
small. The major part of the observed deviations from asymptotic scaling
can be taken care of through a replacement of the bare coupling
by a renormalized coupling \cite{Parisi}. We will use here the definition
\beqn
\beta_{\rm eff} = { 3( N^2-1) \over 2 S_0}~~.
\label{betaeff}
\eqn
With this relation one can determine the cut-off as
\beqn
%a\Lambda_L= R(\beta_{\rm eff})e^{48\pi^2(0.001857-1/352N^2)}
a\Lambda_L= R(\beta_{\rm eff})e^{\pi^2(0.089139-3/(22N^2))}
\label{Reff}
\eqn

\begin{figure}[htb]
\begin{center}
   \epsfig{bbllx=80,bblly=180,bburx=515,bbury=605,
       file=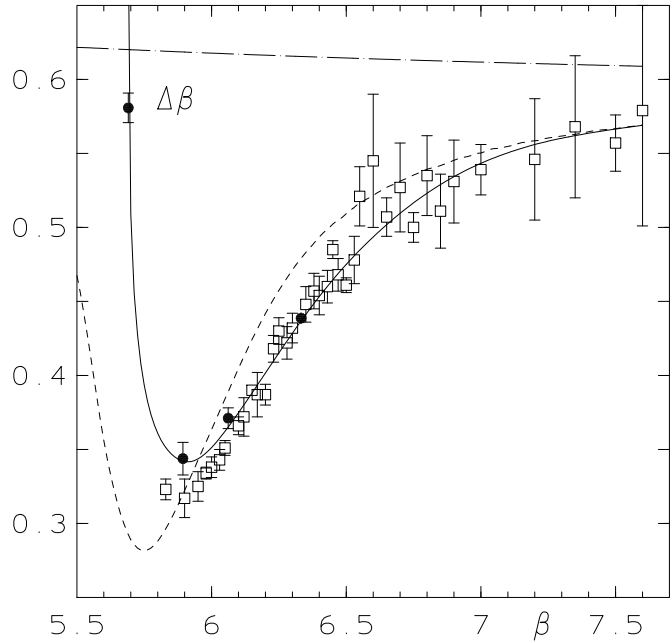, height=90mm}
\end{center}
\caption{The $\Delta\beta$-function,
Eq.~\protect\ref{deltabeta},~from results of
MCRG studies \protect\cite{Akemi} (squares) and from our finite
temperature calculation (dots). The dashed (dashed-dotted) curve shows
$\Delta\beta$ as obtained from the asymptotic form of the
renormalization group equation using the effective (bare) coupling.
The solid curve is derived from our ansatz, Eq.~\protect\ref{scale}. }
\label{fig:deltabeta}
\end{figure}

The deviations from the asymptotic relation given in
Eq.~\ref{renorm} have recently been analyzed for the $SU(3)$ gauge theory
in great detail using a
MCRG analysis of ratios of Wilson loops \cite{Akemi}. Here one calculates
the change in the gauge coupling needed to change the lattice cut-off
by a factor two,
\begin{eqnarray}
\Delta\beta & \equiv & \beta(a) -\beta(2a)
\label{deltabeta}
\end{eqnarray}
The resulting $\Delta\beta$-function is shown in
Figure~\ref{fig:deltabeta} together with $\Delta\beta$ values extracted
from calculations of the critical couplings for the deconfinement
transition, $\Delta\beta = \beta_c(2N_\tau) - \beta_c(N_\tau)$. As can be
seen from this figure, the use of an effective coupling, together with
the asymptotic scaling relation, describes the structure of the observed
scaling violations qualitatively correctly (dashed curve).
Yet, in the region below $\beta \simeq 6.5-7$ there are still considerable
deviations from the data.
In order to get a quantitative description we use therefore the ansatz
\begin{eqnarray}
a\Lambda_L & = & R(\beta) \cdot\lambda (\beta)~~,
\label{scale}
\end{eqnarray}
and determine the function $\lambda (\beta)$ from the measurements of
the critical couplings for $N_\tau = 4,6,8,12$ (Table~\ref{tab:critical})
and for $N_\tau = 3$ from Ref.~\cite{Fin93}. The critical temperature
\beqn
T_c = { 1 \over N_\tau a(\beta_c)}
\label{critte}
\eqn
must be independent of the coupling. The function $\lambda (\beta)$
is then obtained at the critical couplings from
\beqn
\lambda (\beta_c) = { 1 \over N_\tau R(\beta_c) \cdot T_c/\Lambda_L }~~,
\label{critla}
\eqn
up to the unknown constant $T_c/\Lambda_L$. We have fixed this constant
to
\beqn
T_c/\Lambda_L = 34.38~~,
\label{crittev}
\eqn
by first interpolating $1/N_\tau R(\beta_c)$ in $\beta$ and then
requiring a smooth transition of the resulting $\Delta\beta-$function
into that of the effective coupling scheme at higher $\beta-$values.
%such that the existing numerical results for $\Delta\beta$ are reproduced.
This is shown as solid curve in Figure~\ref{fig:deltabeta}.
Quite obviously, both the critical couplings and $\Delta\beta$ are reproduced
above $\beta \approx 6.0$ very well by the introduction of one correction
function $\lambda (\beta)$. At smaller $\beta-$values, however, we observe 
differences.

In Table~\ref{tab:scale} we give our results for the function
$\lambda(\beta)$, which allows a direct evaluation of the
lattice cut-off $a\Lambda_L$ from Eq.~\ref{scale}.
Knowing the cut-off as a function of
the gauge coupling $\beta$ we can also evaluate the derivative
$ad\beta / da$, which is needed for the calculation of
the difference $\epsilon-3p$ (Eq.~\ref{deltalat}).
The corresponding values are also given in Table~\ref{tab:scale}.
%\vskip 1truecm
\begin{table}
\begin{center}
\begin{tabular}{|c|c|c|c|c|}
\hline
\multicolumn{1}{|c|}{$\beta=2N/g^{2}$} &
\multicolumn{1}{|c|}{$\lambda$} &
\multicolumn{1}{|c|}{$a dg^{-2} / da$} &
\multicolumn{1}{|c|}{$\partial g^{-2}_{\sigma} / \partial \xi$} &
\multicolumn{1}{|c|}{$\partial g^{-2}_{\tau} / \partial \xi$} \\
\hline
    5.70   &   2.194700   &   -0.077855   &    0.378981   &   -0.340054   \\
    5.75   &   2.087578   &   -0.079269   &    0.372659   &   -0.333024   \\
    5.80   &   1.990720   &   -0.081694   &    0.366500   &   -0.325653   \\
    5.85   &   1.905342   &   -0.085053   &    0.360557   &   -0.318030   \\
    5.90   &   1.831481   &   -0.089150   &    0.354868   &   -0.310293   \\
    5.95   &   1.768642   &   -0.093699   &    0.349471   &   -0.302621   \\
    6.00   &   1.715383   &   -0.098172   &    0.344400   &   -0.295314   \\
    6.05   &   1.670045   &   -0.102343   &    0.339672   &   -0.288501   \\
    6.10   &   1.631051   &   -0.105965   &    0.335221   &   -0.282238   \\
    6.15   &   1.596995   &   -0.109147   &    0.330959   &   -0.276386   \\
    6.20   &   1.567082   &   -0.112127   &    0.326800   &   -0.270737   \\
    6.25   &   1.540692   &   -0.114874   &    0.322688   &   -0.265252   \\
    6.30   &   1.517309   &   -0.117373   &    0.318698   &   -0.260011   \\
    6.35   &   1.496477   &   -0.119668   &    0.314933   &   -0.255099   \\
    6.40   &   1.477892   &   -0.121832   &    0.311500   &   -0.250584   \\
    6.45   &   1.461275   &   -0.123799   &    0.308467   &   -0.246568   \\
    6.50   &   1.446329   &   -0.125501   &    0.305763   &   -0.243012   \\
    6.55   &   1.432777   &   -0.127015   &    0.303276   &   -0.239769   \\
    6.60   &   1.420468   &   -0.128442   &    0.300900   &   -0.236679   \\
    6.65   &   1.409274   &   -0.129750   &    0.298556   &   -0.233681   \\
    6.70   &   1.399057   &   -0.130908   &    0.296289   &   -0.230835   \\
    6.75   &   1.389672   &   -0.131888   &    0.294178   &   -0.228235   \\
    6.80   &   1.380963   &   -0.132664   &    0.292300   &   -0.225968   \\
    6.85   &   1.372792   &   -0.133297   &    0.290707   &   -0.224058   \\
    6.90   &   1.365080   &   -0.133855   &    0.289355   &   -0.222428   \\
    6.95   &   1.357773   &   -0.134348   &    0.288176   &   -0.221001   \\
    7.00   &   1.350826   &   -0.134792   &    0.287100   &   -0.219704   \\
    7.05   &   1.344204   &   -0.135196   &    0.286069   &   -0.218471   \\
    7.10   &   1.337881   &   -0.135574   &    0.285065   &   -0.217278   \\
    7.15   &   1.331835   &   -0.135921   &    0.284078   &   -0.216117   \\
    7.20   &   1.326041   &   -0.136234   &    0.283100   &   -0.214983   \\
 $\infty$  &   1.000000   &   -0.139316   &    0.201600   &   -0.131940   \\
\hline
\end{tabular}
\end{center}
\caption{The correction factor $\lambda$ and non-perturbative
coupling derivatives.}
\label{tab:scale}
\end{table}

%%%%%%%%%%%%%%%%%%%%%%%%%%%%%%%%%%%%%%%%%%%%%%%%%%%%%%%%%%%%%%%%%%%%%%%%%%%%%%%

\section{Thermodynamics on Finite Lattices: $N_\tau$=4, 6 and 8}

%%%%%%%%%%%%%%%%%%%%%%%%%%%%%%%%%%%%%%%%%%%%%%%%%%%%%%%%%%%%%%%%%%%%%%%%%%%%%%%

\subsection{Perturbative High Temperature Limit}

%%%%%%%%%%%%%%%%%%%%%%%%%%%%%%%%%%%%%%%%%%%%%%%%%%%%%%%%%%%%%%%%%%%%%%%%%%%%%%%
In order to extract physical observables in the continuum limit from
results on finite lattices at finite lattice cut-off one has to
control the infrared as well as ultraviolet effects
that influence them. In the case of thermodynamic observables
infrared and ultraviolet effects are of importance in different
temperature regimes. Close to the phase transition low
momentum modes are most dominant and the finite physical volume thus
influences physical observables strongly. In this regime
thermodynamic quantities strongly depend on $TV^{1/3}\equiv N_\sigma /
N_\tau$. This has been shown for the $SU(2)$ theory
\cite{Eng95} and was used there to extract the critical energy
density in the thermodynamic limit.

At high temperature, high momentum modes
($|\vec{p}| \sim T$) give the largest contribution to the energy density
and pressure, at least to the extent that the momentum spectrum is well 
approximated by that of an ideal gas. This contribution is therefore strongly
influenced by the finite lattice cut-off. The standard one-plaquette
Wilson action used in lattice calculations provides a discrete version of
the field strength tensor, which leads to deviations of
$O(a^2)$ of the lattice Lagrangian from its continuum counterpart.
Thermodynamic observables like energy density and pressure will then deviate
from the continuum values by $O((aT)^2)$ terms (note that $aT\equiv 1/N_\tau$).
For a non-interacting gluon gas, which corresponds to the $g^2 \rightarrow
0$ limit of the finite temperature $SU(N)$ gauge theory, one finds
explicitly \cite{Eng95,Bei95}
\beqn
{\epsilon\over T^4} = {3p \over T^4} =
(N^2-1) {\pi^2 \over 15}\biggl[ 1+
 {30  \over 63} \cdot \biggl ({\pi \over N_\tau} \biggr )^2 +
 {1 \over 3} \cdot \biggl ({\pi \over N_\tau } \biggr )^4 +
O\biggl ({1 \over N_\tau^6} \biggr ) \biggr ]~.
\label{Ntdependence}
\eqn
Apparently the subleading corrections are still large for
lattices with small temporal extent like $N_\tau=4$. We therefore will
restrict our extrapolations to the continuum limit
($N_\tau \rightarrow \infty$) to the numerical
results obtained on lattices with temporal extent $N_\tau =6$ and 8.
We will work on large spatial lattices, $N_\sigma / N_\tau = (4-5.33)$,
which will ensure that we are close to the thermodynamic limit
for a wide temperature range except very close to $T_c$.

%%%%%%%%%%%%%%%%%%%%%%%%%%%%%%%%%%%%%%%%%%%%%%%%%%%%%%%%%%%%%%%%%%%%%%%%%%%%%%

\subsection{Monte Carlo Results}

%%%%%%%%%%%%%%%%%%%%%%%%%%%%%%%%%%%%%%%%%%%%%%%%%%%%%%%%%%%%%%%%%%%%%%%%%%%%%%
As explained in section 2, the basic observable entering the calculations
of the pressure is the difference between the expectation values of the
action densities calculated at zero temperature ($N_\tau = N_\sigma$) and
at finite temperature ($N_\tau < N_\sigma$). For dimensional reasons these
differences will scale like  $N_\tau^{-4}$. We thus consider the quantities
\beqn
\Delta S = N_\tau^4 \bigl( S_0 - S_T\bigr)~.
\label{plaqdif}
\eqn
Due to the strong $N_\tau$-dependence of $\Delta S$ a high accuracy of
the plaquette expectation values is
needed.  We have calculated plaquette expectation values on lattices of size
$16^4$, $32^4$ as well as $16^3\times 4$, $32^3\times 6$ and
$32^3\times 8$ for a large number of gauge couplings. A collection
of all our results as well as further information about the actual Monte
Carlo calculation is given in the Appendix.

\begin{figure}[htb]
\begin{center}
  \epsfig{bbllx=45,bblly=50,bburx=480,bbury=670,
       file=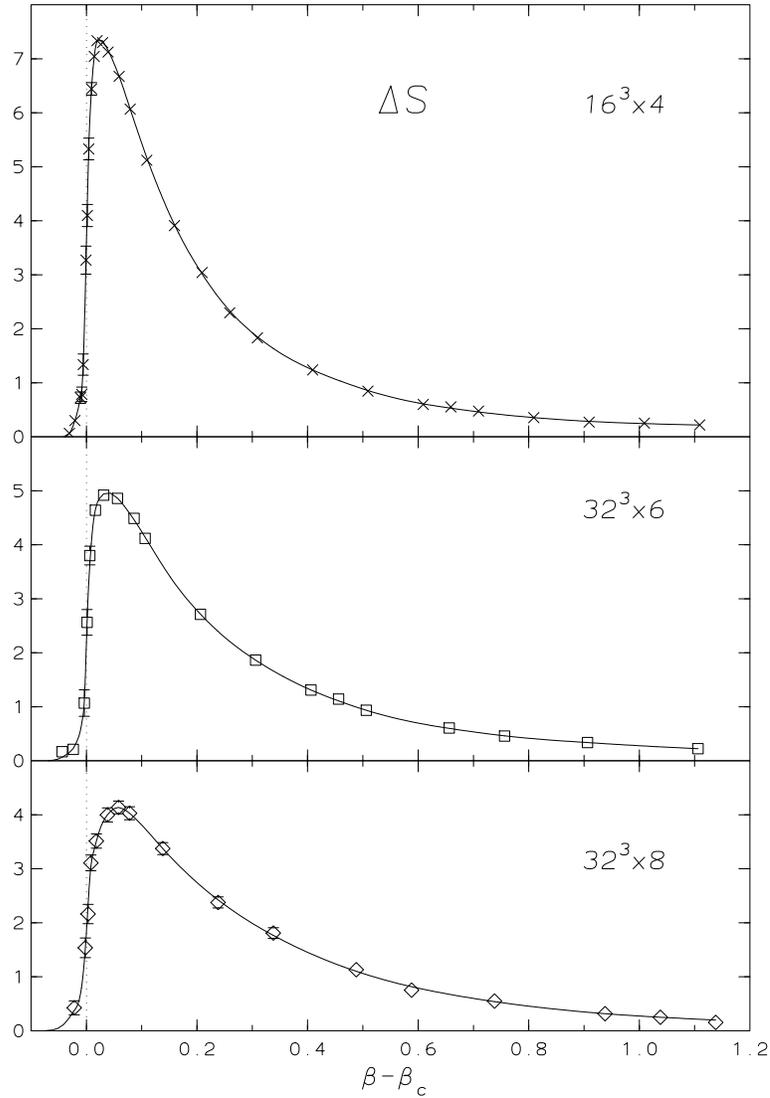, height=130mm}
\end{center}
\caption{The difference $\Delta S$ versus
$\beta- \beta_c (N_\tau, N_\sigma)$ for $N_\tau = 4$, 6 and 8. }
\label{fig:plaqdif}
\end{figure}

In Figure~\ref{fig:plaqdif} we show the differences $\Delta S$
for $N_\tau =4$, 6 and 8.
As can be seen their magnitude remains $O(1)$ with increasing
$N_\tau$. Note, however, a drop
of the maximum of the plaquette difference with increasing $N_\tau$ as well
as a widening of the peak as function of $\beta$. Both effects reflect the
presence of finite cut-off effects, which will show up in the pressure
as well as the energy density calculated from $\Delta S$.

For a calculation of the pressure we have to integrate $\Delta S$ with
respect to $\beta$. In order to estimate the error resulting from this
integration we performed it in two different ways. The most straightforward
way is to use a straight line interpolation between neighbouring data
points. We may, however, make use of the fact that the values
of $\Delta S$ calculated at different values of $\beta$ are statistically
independent and use a (spline) interpolating curve to smoothen statistical
fluctuations.  These smooth interpolations are shown in Fig.~\ref{fig:plaqdif}.
We also see in this figure  that $\Delta S$  rapidly becomes small below
the critical coupling $\beta_c(N_\tau, N_\sigma)$. We thus can use a
value $\beta_0$ close to $\beta_c$ for the
normalization of the free energy density and calculate the pressure by
integrating the plaquette differences as given in Eq.~\ref{freelat}.

In order to represent the pressure in terms of $T/T_c$ we will use
the two parametrizations (Eqs.~\ref{Reff} and \ref{scale}) discussed in the
previous section. This will give us some idea about the systematic errors
induced through such a parametrization. In the case of the pressure a change
in the parametrization will only amount to a shift in the temperature scale.
This effect is largest for small values of $N_\tau$.

\begin{figure}[htb]
\begin{center}
  \epsfig{bbllx=60,bblly=170,bburx=550,bbury=640,
       file=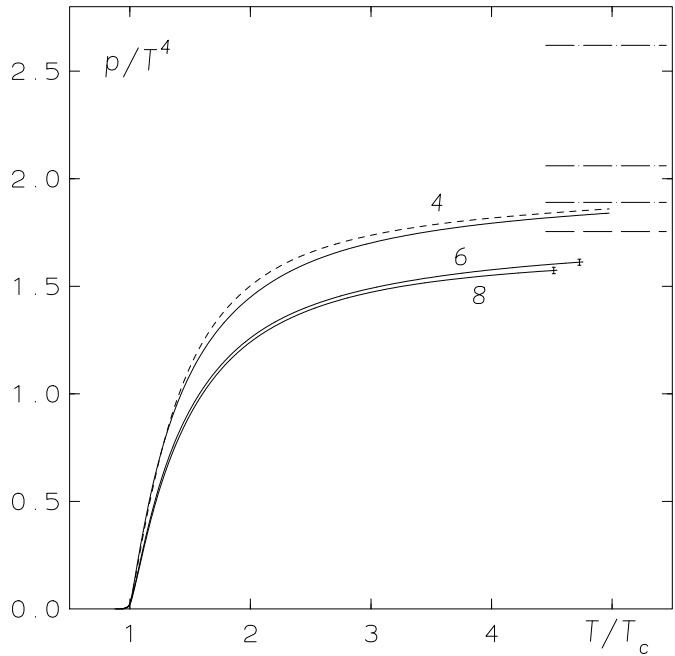, height=90mm}
\end{center}
\caption{The pressure versus $T/T_c$ for $N_\tau = 4$, 6 and 8
integrating
the interpolations for the action density. For $N_\tau = 4$ we show two
curves, one for the
temperature scale using the effective coupling scheme (dashed curve) and one
for our parametrization
(solid curve). For $N_\tau = 6,$ and 8 we only show the
latter.
The error bars indicate the uncertainties arising from the integration.
The horizontal dashed line shows the ideal gas continuum value, the
dashed-dotted lines the corresponding lattice values for $N_\tau =4$, 6 and 8.}
\label{fig:pressure}
\end{figure}

The results for the pressure as a function of $T/T_c$ obtained from
calculations with temporal extent $N_\tau =4$, 6 and 8 are shown in
Figure~\ref{fig:pressure}. We clearly see the expected finite size dependence.
It qualitatively reflects the $N_\tau$-dependence of the free gluon gas, which
is shown by horizontal dashed-dotted lines in this figure.
Quantitatively, however, we find that the cut-off dependence of the pressure
is considerably weaker than suggested by the free gas calculation. We will
discuss this in detail in the next section.

\begin{figure}[htb]
\begin{center}
 \epsfig{bbllx=75,bblly=165,bburx=540,bbury=645,
       file=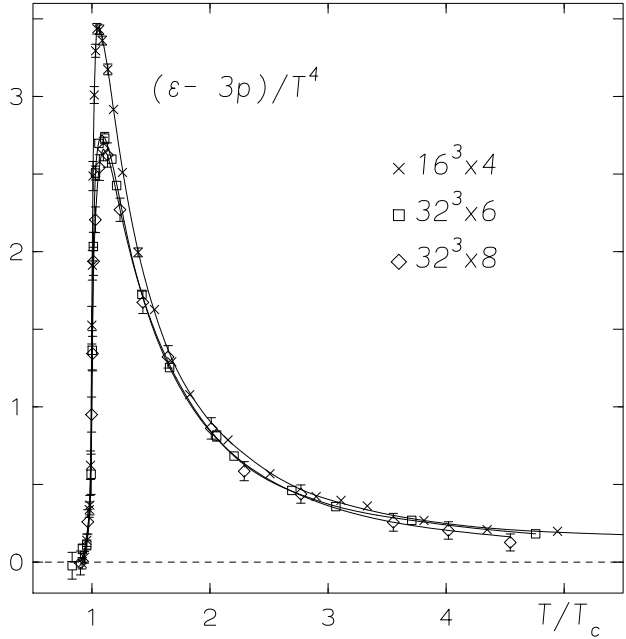, height=90mm}
\end{center}
\caption{The difference $(\epsilon -3p)/T^4$.}
\label{fig:e3p}
\end{figure}

Unlike the pressure, the absolute magnitude of the energy density
depends on the derivative
$Td\beta / dT$. Again we will observe the strongest
sensitivity to the parametrization of the temperature scale in the $N_\tau =4$
case, where the numerical calculations close to $T_c$ are performed with bare
couplings $\beta \in [5.6,5.8]$, \ie~in the region of strongest variation of
the cut-off with $\beta$ (dip in the $\Delta \beta$-function).
In Fig.~\ref{fig:e3p} we show the difference $(\epsilon - 3p)/T^4$ obtained
with the parametrization of the temperature scale specified in Eq.~\ref{scale}.
For $N_\tau=6$ and 8 the influence of the parametrization is not
significant and is well inside the error bars of the data.

The first order nature of the $SU(3)$ phase transition has clearly been
established in the studies on lattices with $N_\tau = 4$ and 6 and the
latent heat at $T_c$ has been estimated in these cases \cite{Iwa92}. Also in
our calculations for $N_\tau =8$ and 12 we see clear metastabilities (phase
flips of the Polyakov loop) at the critical coupling. Our
statistics were, however, not sufficient to extract the latent heat from the
discontinuity in the plaquette expectation values. We thus did not attempt
to separate our data sample in the vicinity of $\beta_c$ into sets belonging
to different phases. Consequently our interpolation curves for $(\epsilon
-3p)/T^4$, shown in Fig.~\ref{fig:e3p}, are continuous (as it should be for
calculations performed in finite physical volumes).

\begin{figure}[htb]
\begin{center}
  \epsfig{bbllx=45,bblly=165,bburx=550,bbury=635,
       file=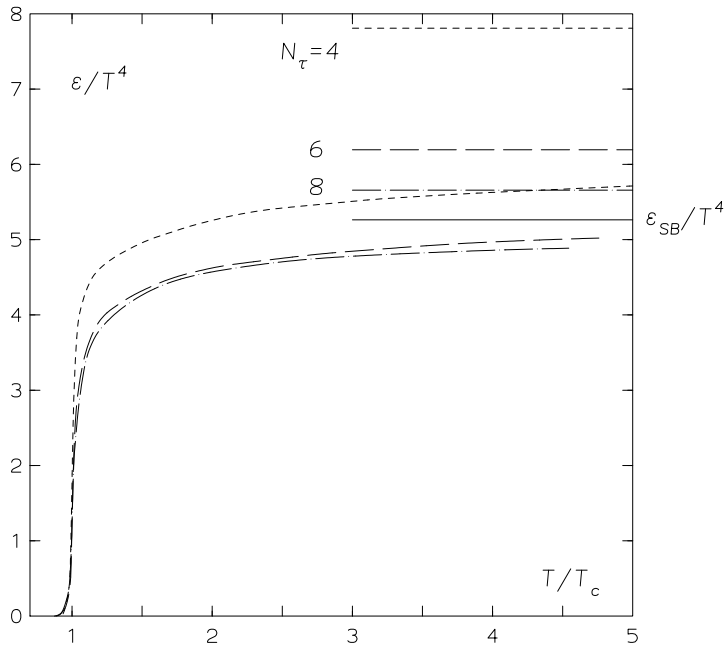, height=90mm}
\end{center}
\caption{The energy density versus $T/T_c$ for $N_\tau = 4$
(short dashes), 6 (dash-dotted line) and 8 (long dashes).}
\label{fig:energy}
\end{figure}

Combining the results for the pressure and the difference $\epsilon -3p$
we can calculate the energy density. This is shown in Fig.~\ref{fig:energy}
for the three different lattice sizes.

%%%%%%%%%%%%%%%%%%%%%%%%%%%%%%%%%%%%%%%%%%%%%%%%%%%%%%%%%%%%%%%%%%%%%%%%%%%%%%

\section{Thermodynamics in the Continuum Limit}

%%%%%%%%%%%%%%%%%%%%%%%%%%%%%%%%%%%%%%%%%%%%%%%%%%%%%%%%%%%%%%%%%%%%%%%%%%%%%%
Based on the analysis of the pressure and energy density on various size
lattices, which we have presented in the previous section, we can
attempt now to extrapolate these quantities to the continuum limit. As
discussed in the previous section, the leading corrections to the
continuum limit are $O(N_\tau^{-2})$. The analysis of the high temperature
($g^2 = 0$) ideal gas limit, however, shows that for $N_\tau =4$ higher
order corrections are expected to be very important. We thus will perform an
extrapolation to the continuum limit only on the basis of our numerical
results for $N_\tau =6$ and 8.
We have performed extrapolations using the ansatz,
\beqn
\biggl({p \over T^4}\biggr)_a = \biggl({p \over T^4}\biggr)_0
+ {c_2(T) \over N_\tau^2} ~~,
\label{cfit}
\eqn
We generally find that the difference between the extrapolated values and
the results for $N_\tau=8$ is less than 3\%, which should be compared with
the corresponding result for the free gas, where the difference is still
about 10\%. At our highest temperatures, $T/T_c \simeq 4-5$,
the fit parameter $c_2$ takes values which are only half as large as the
expansion parameter for an ideal gas (Eq.~\ref{Ntdependence}).

\begin{figure}[htb]
\begin{center}
  \epsfig{bbllx=55,bblly=65,bburx=535,bbury=730,
       file=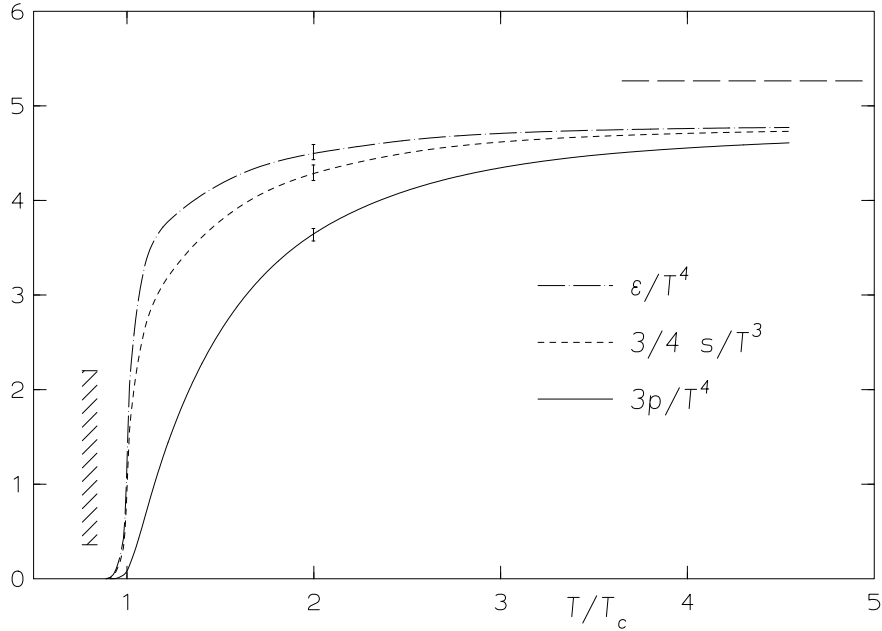, height=120mm, angle=-90}
\end{center}
\caption{Extrapolation to the continuum limit for the energy density,
entropy density and pressure versus $T/T_c$. The dashed horizontal line
shows the ideal gas limit. The hatched vertical band indicates the size of the 
discontinuity in $\epsilon/T^4$ (latent heat) at $T_c$ \protect\cite{Iwa92}.
Typical error bars are shown for all curves.}
\label{fig:continuum}
\end{figure}

The cut-off dependence in $(\epsilon -3p)/T^4$ is much smaller than for
the pressure alone. Within errors the interpolation
curves shown for $N_\tau=6$ and 8 in Fig.~\ref{fig:e3p} are already coinciding.
Combining therefore
$(\epsilon -3p)/T^4$ for $N_\tau=8$ and the extrapolations for the pressure
we obtain the
continuum results for other thermodynamic quantities like the energy
density and entropy density, $s=(\epsilon +p)/T$. These results are
summarized in Fig.~\ref{fig:continuum}.

%%%%%%%%%%%%%%%%%%%%%%%%%%%%%%%%%%%%%%%%%%%%%%%%%%%%%%%%%%%%%%%%%%%%%%%%%%%%%%

\section{Discussion}

%%%%%%%%%%%%%%%%%%%%%%%%%%%%%%%%%%%%%%%%%%%%%%%%%%%%%%%%%%%%%%%%%%%%%%%%%%%%%%

\subsection{The Pressure Revisited}

%%%%%%%%%%%%%%%%%%%%%%%%%%%%%%%%%%%%%%%%%%%%%%%%%%%%%%%%%%%%%%%%%%%%%%%%%%%%%%
Our present determination of thermodynamic quantities has been based
on a calculation of the free energy density through the integration of
plaquette expectation values (Eq.~\ref{freelat}). This approach has
been introduced some time ago in order to avoid the use of perturbative
relations for the derivatives of coupling constants with respect to
the lattice cut-off in spatial ($a$) and temporal ($a_\tau$)
directions \cite{Eng90,Kar82}, which appear when one calculates pressure
and energy density from derivatives of the free energy with respect to volume
and temperature. The pressure, for instance, is obtained as
\begin{eqnarray}
 { p \over T^4}&=&
 NN_\tau^4\biggl\{
             \biggl[2g^{-2}-
\biggl({{\partial} g_\sigma^{-2} \over{\partial}\xi}
- {{\partial}  g_\tau^{-2} \over {\partial} \xi}\biggr)\biggr]
  (P_{\sigma}-P_{\tau})
\nonumber   \\
& &\qquad -3 \biggl({{\partial} g_\sigma^{-2} \over{\partial}\xi}
+ {{\partial}  g_\tau^{-2} \over {\partial} \xi}\biggr)
  \biggl[ 2P_0-(P_{\sigma}+P_{\tau}) \biggr]
       \biggr\}~,
\label{dpressure}
\end{eqnarray}
where $g_\sigma$ and $g_\tau$ denote the gauge couplings
in front of the space-space and space-time plaquettes in the action,
$\xi= a/a_\tau$ and the derivatives are taken at $\xi =1$.\footnote{We
follow here closely the discussion of the $SU(2)$ case presented in
\cite{Eng95}.}
A comparison of this relation with our new results for the pressure
yields one equation for the derivatives of the coupling constants.
A second equation is obtained from the well-known connection of the
$\beta-$function to the sum of derivatives
\beqn
a{dg^{-2} \over da} =
-2 \biggl({{\partial} g_\sigma^{-2} \over{\partial}\xi}
+ {{\partial}  g_\tau^{-2} \over {\partial} \xi}\biggr)_{\xi=1}~~,
\label{sumrule}
\eqn
so that both derivatives may be calculated non-perturbatively.
The numerical determination of the derivatives below the transition
point, where both the pressure and the plaquette difference $P_{\sigma}-
P_{\tau}$ are becoming small, will, however, lead to increasing errors.
In addition, we expect finite size effects because the transition is of
first order in the thermodynamic limit, but continuous on finite lattices.
Indeed, we find comparable results above the transition points for
$N_{\tau}=4,6$ and 8, but deviations in the close vicinity of each
critical $\beta-$value. In Table~\ref{tab:scale} we list the results from
an interpolation of our directly calculated derivative values. As for
$SU(2)$ we observe strong deviations from the asymptotic values \cite{Kar82}
up to large coupling constant values.

\begin{figure}[htb]
\begin{center}
 \epsfig{bbllx=70,bblly=170,bburx=545,bbury=625,
       file=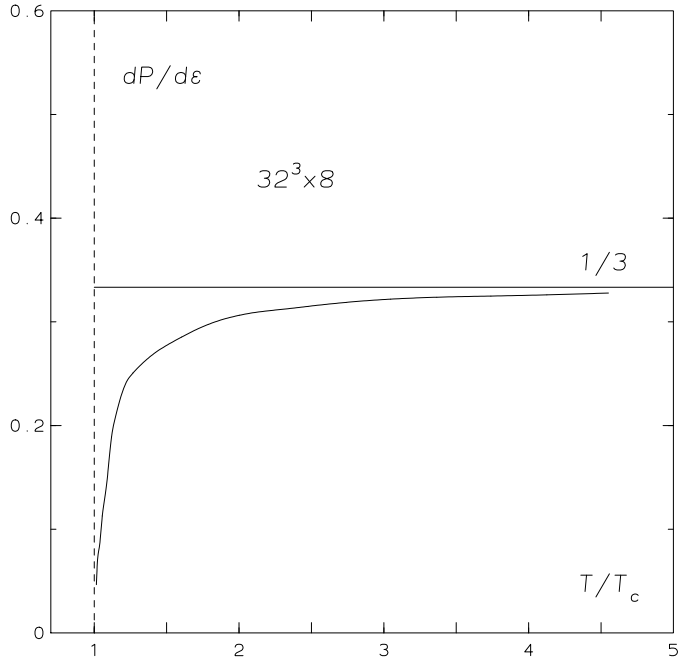, height=90mm}
\end{center}
\caption{The square of the velocity of sound versus $T/T_c$ for
$SU(3)$.}
\label{fig:sound}
\end{figure}

At this point the following remark may be appropriate.
The derivatives of the coupling constants are also of importance at
the transition point, where they influence the relative size of the
gaps in the space--space and space--time plaquettes. Since the pressure
is continuous also at the first order transition point, we find
from Eq.~\ref{dpressure} for the differences $\Delta P_{\sigma,\tau}$
between the plaquettes just above and below the transition the
following relation
\beqn
\biggl[ g^{-2} -a {dg^{-2} \over da} - {\partial g^{-2}_{\sigma} \over
\partial \xi} \biggr] \Delta P_{\sigma} =
\biggl[ g^{-2} +a {dg^{-2} \over da} + {\partial g^{-2}_{\tau} \over
\partial \xi} \biggr] \Delta P_{\tau}~~.
\label{gap}
\eqn
As a consequence, the gaps in $P_{\sigma}$ and $P_{\tau}$ will be of
the same size for large $N_\tau$, \ie~in the weak coupling region
\beqn
{\Delta P_{\sigma} \over \Delta P_{\tau}} = 1 ~~~~{\rm for}~~N_{\tau}
\rightarrow
\infty~,
\label{gaplimit}
\eqn
because the $g^{-2}-$terms will dominate over the constant derivatives.
%%%%%%%%%%%%%%%%%%%%%%%%%%%%%%%%%%%%%%%%%%%%%%%%%%%%%%%%%%%%%%%%%%%%%%%%%%%%%%

\subsection{The Velocity of Sound}

%%%%%%%%%%%%%%%%%%%%%%%%%%%%%%%%%%%%%%%%%%%%%%%%%%%%%%%%%%%%%%%%%%%%%%%%%%%%%%
A quantity of much interest for the hydrodynamic description of the
evolution of a quark gluon plasma, which eventually will get created
in ultra-relativistic heavy ion collisions, is the velocity of sound, $c_s$.
It is obtained as
\beqn
c_s^2 = {{\rm d} p \over {{\rm d} \epsilon}} ~~.
\label{sound}
\eqn
We can determine this quantity easily from the continuum extrapolations
of the energy density and the pressure. The result is shown in
Fig.~\ref{fig:sound}. We note the rapid decrease of $c_s$ in the vicinity
of the deconfinement phase transition, which reflects the rapid rise of
the energy density at $T_c$ and the small latent heat at the transition
temperature. The transition is {\it almost second order} in this respect.

%%%%%%%%%%%%%%%%%%%%%%%%%%%%%%%%%%%%%%%%%%%%%%%%%%%%%%%%%%%%%%%%%%%%%%%%%%%%%%

\subsection{The Gluon Condensate}

%%%%%%%%%%%%%%%%%%%%%%%%%%%%%%%%%%%%%%%%%%%%%%%%%%%%%%%%%%%%%%%%%%%%%%%%%%%%%%
After a rapid rise close to $T_c$ all normalized 
thermodynamic observables (\ie~divided by $T^4$)
show a rather slow increase at temperatures above $(2-3)T_c$. This
suggests that they may be described in terms of functions of a
coupling constant $g^2(T)$, which only depends logarithmically on
the temperature due to the QCD renormalization group equation. However, it
also is clear that the known perturbative expansion of the QCD free energy
density \cite{Arnold,Zhai} is not adequate to describe the numerical
results quantitatively. A quantity which is very sensitive to
deviations from the perturbative high temperature limit
is the trace anomaly of the energy-momentum tensor,
$T^\mu_\mu \equiv \epsilon - 3p$. It is connected to the temperature
dependent gluon condensate \cite{Leu92} via
\beqn
\epsilon -3p = \langle G^2 \rangle_{T=0} - \langle G^2 \rangle_T ~~.
\label{condensate}
\eqn
At high temperature the left hand side of this equation has been calculated
in continuum perturbation theory \cite{Kap79}
\beqn
{\epsilon -3p \over T^4} = \tilde{\beta}(g)
\biggl[{(N^2-1) \over 288}  g^4(T) -  {N^2-1 \over 16N\pi} \biggl(
{N\over 3}\biggr)^{3/2}  g^5 (T)
+ O(g^6) \biggr]~~,
\label{pertcond}
\eqn
where $\tilde{\beta}$ is defined in Eq.~\ref{rgeqn} and its two first
perturbative terms are given by,
\beqn
\tilde{\beta}(g) = 4N [b_0+b_1 g^2] ~~.
\label{pertrge}
\eqn

\begin{figure}[htb]
\begin{center}
 \epsfig{bbllx=75,bblly=160,bburx=540,bbury=640,
       file=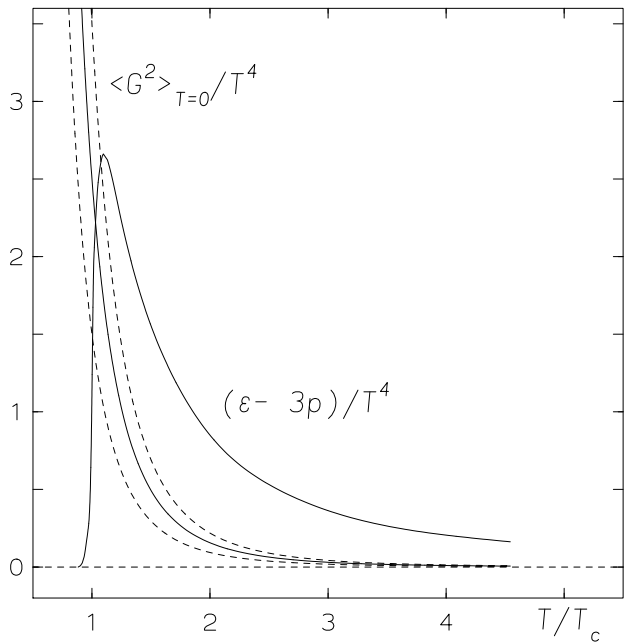, height=90mm}
\end{center}
\caption{The zero temperature gluon condensate divided by $T^4$
and $(\epsilon -3p)/T^4$ versus $T/T_c$. The dashed lines show the 
uncertainty of the zero temperature gluon condensate value.}
\label{fig:condensate}
\end{figure}

This shows that $\langle G^2 \rangle_T$ has to be
negative and proportional to $g^4(T) T^4$ at high temperature. The first
term on the right hand side, $ \langle G^2 \rangle_{T=0}$, thus will soon
be negligible compared to the second term.

We may re-interpret the numerical results for $\epsilon -3p$ in terms of
the temperature dependence of the gluon condensate. To do so we have to
subtract from the numerical results the zero temperature gluon
condensate. We use $ \langle G^2 \rangle_{T=0} \simeq (1-2)~$GeV/fm$^3$
\cite{Leu92}, where we allow for quite a large uncertainty.
This can be expressed in units of $T_c$,
\beqn
 \langle G^2 \rangle_{T=0} \simeq (2.5 \pm 1.0)~T_c^4~~.
\label{condensateTc}
\eqn
The result for $ \langle G^2 \rangle_{T=0} /T^4$
is shown in Fig.~\ref{fig:condensate} together with $(\epsilon
-3p)/T^4$. As can be seen, the
exact value of the zero temperature condensate is not important for the
high temperature behaviour. 
Close to $T_c$ the difference $\epsilon - 3p$ is essentially given by
$\langle G^2 \rangle_{T=0}$, i.e. the finite temperature condensate 
$\langle G^2 \rangle_{T}$, which shows little temperature dependence below
$T_c$, decreases rapidly above $T_c$ and yields the dominant (negative) 
contribution for $T\gsim 2~T_c$. Fig.~\ref{fig:condensate} shows
that the contribution of the zero temperature condensate rapidly becomes
negligible above $T_c$.
Although this comparison shows that the rise of
$(\epsilon -3p)/T^4$ close to $T_c$ is related to the non-perturbative
vacuum contributions to the equation of state, it also makes clear that
this is not sufficient to explain the large values for this quantity
at temperatures $T\ge 2T_c$. This also cannot easily be explained by
the perturbative relation given in Eq.~\ref{pertcond}.
If one insists, however, on a comparison with the leading perturbative
term only, one has to conclude that the temperature dependent
running coupling has to be large, $g^2(T) \simeq 2$ even at $T \simeq 5
T_c$.
This is consistent with our analysis of the electric and magnetic
condensates as well as the
spatial string tension, which are discussed in the next two subsections.
 
%%%%%%%%%%%%%%%%%%%%%%%%%%%%%%%%%%%%%%%%%%%%%%%%%%%%%%%%%%%%%%%%%%%%%%%%%%%%%%

\subsection{Magnetic and Electric Condensates}

%%%%%%%%%%%%%%%%%%%%%%%%%%%%%%%%%%%%%%%%%%%%%%%%%%%%%%%%%%%%%%%%%%%%%%%%%%%%%%
The difference $(\epsilon - 3p)/T^4$, as well as
$\epsilon /T^4$ and $p/T^4$
separately, are given by specific combinations of
spacelike and timelike plaquette expectation values. The
spacelike plaquettes $P_\sigma$ are related to
the spatial components of $F^2_{i j}$, $i,j =1,2,3$,
\ie~the square of the magnetic field strength. Similarly, the timelike
plaquettes are related to the square of the (Euclidean) electric field
strength, $F^2_{0,i}$.
The different components have been used as an indicator for the behaviour of
the magnetic and electric parts of the gluon condensate at high temperature
\cite{emcondens}. It therefore is of interest to determine
the contributions of space and timelike plaquette
expectation values to the difference $\epsilon-3p$ separately,
\begin{eqnarray}
\Delta_\sigma \equiv 3N_\tau^4 \tilde{\beta}[ P_0 - P_\sigma ]~~,
\nonumber \\
\Delta_\tau \equiv 3N_\tau^4 \tilde{\beta}[ P_0 - P_\tau ]~~.
\label{deltaem}
\end{eqnarray}
As can be seen
from the perturbative result for the plaquettes \cite{Hel85},
the leading order contributions to
$\Delta_{\sigma(\tau)}$ are $O(g^2)$, of opposite sign and equal size

\begin{eqnarray}
\Delta_\tau = -\Delta_\sigma &=& N_\tau^4 b_0 {N^2-1 \over 2} g^2
\bigl(1 - 4 I_4 \bigr) + O(g^4)~~,
\label{pertDelta}
\end{eqnarray}
with
\begin{eqnarray}
I_4 &=& {1\over N^3_\sigma N_\tau} \sum_{n_\mu} {\sin^2 \bigl(\pi
n_4/N_\tau \bigr) \over \sum_{i=1}^3 \sin^2 \bigl(\pi n_i /N_\sigma
\bigr) + \sin^2 \bigl( \pi n_4 /N_\tau \bigr) } \quad ,
\label{Ifour}
\end{eqnarray}
where $n_i = 0,$ 1, ...$N_\sigma -1$ and
$n_4 = 0,$ 1, ...$N_\tau -1$. While $\Delta_\tau$ and $\Delta_\sigma$
are $O(g^2)$, these leading order contributions, including all $O(a^n
g^2)$ cut-off effects, cancel in
$(\epsilon - 3p)/T^4 = \Delta_\sigma + \Delta_\tau$, which is $O(g^4)$.
The perturbative expansion of $\Delta_\tau$ is closely related to that
of the energy density. In the continuum limit we find
\begin{eqnarray}
\Delta_\tau = {11 N \over 720} \bigl( N^2 -1 \bigr) g^2 + O(g^4)~~.
\label{deltac}
\end{eqnarray}

\begin{figure}[htb]
\begin{center}
 \epsfig{bbllx=70,bblly=165,bburx=545,bbury=635,
       file=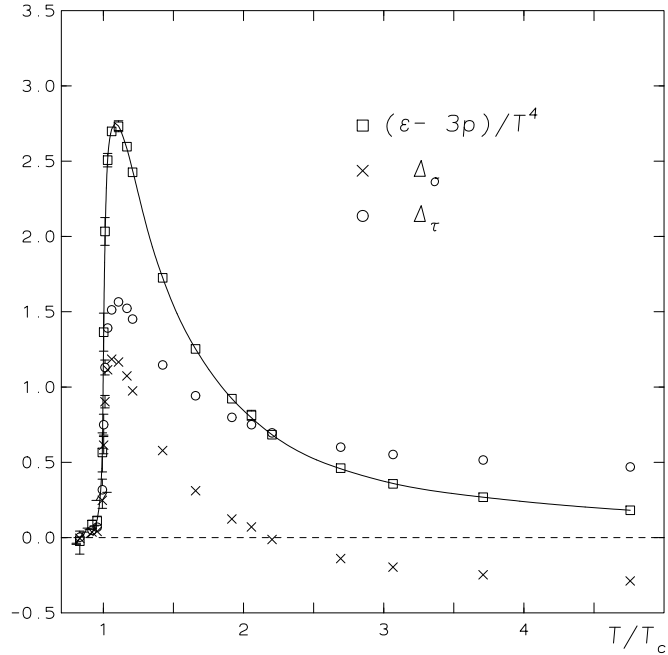, height=90mm}
\end{center}
\caption{The space- and timelike components
$\protect\Delta_{\sigma}$ and $\protect\Delta_{\tau}$ of
 $(\epsilon-3p)/T^4$ versus $T/T_c$ on a $32^3 \times 6$ lattice.}
\label{fig:emcondens}
\end{figure}

We show $\Delta_\sigma$ and $\Delta_\tau$ in Fig.~\ref{fig:emcondens}. We
note that up to $T_{\rm peak}\simeq 1.1T_c$ both terms contribute roughly the 
same amount to $\epsilon -3p$. Furthermore, the magnitudes of both quantities
remain large even at $5T_c$. In fact, a comparison with Eq.~\ref{deltac}
shows that $g^2(5T_c) \simeq 1.4$ is needed in order to describe
$\Delta_\sigma$ and $\Delta_\tau$ in terms of the leading order
perturbative relation.

%%%%%%%%%%%%%%%%%%%%%%%%%%%%%%%%%%%%%%%%%%%%%%%%%%%%%%%%%%%%%%%%%%%%%%%%%%%%%%

\subsection{Spatial String Tension}

%%%%%%%%%%%%%%%%%%%%%%%%%%%%%%%%%%%%%%%%%%%%%%%%%%%%%%%%%%%%%%%%%%%%%%%%%%%%%%

The spatial string tension provides some insight into the structure of the
effective three dimensional gauge-Higgs theory, which can be constructed
by systematically integrating out heavy modes.

It has been analyzed recently in quite some
detail for the $SU(2)$ gauge theory \cite{sst2}. First results for $SU(3)$
have been reported in \cite{sst3}. The spatial string tension, $\sigma_s$,
is extracted from the asymptotic behaviour of large spatial Wilson
loops, which show an area law behaviour even above $T_c$
\beqn
\langle W(R,S) \rangle \sim e^{-\sigma_s RS} ~~.
\label{sigmas}
\eqn
\begin{figure}[htb]
\begin{center}
 \epsfig{bbllx=55,bblly=55,bburx=540,bbury=740,
         file=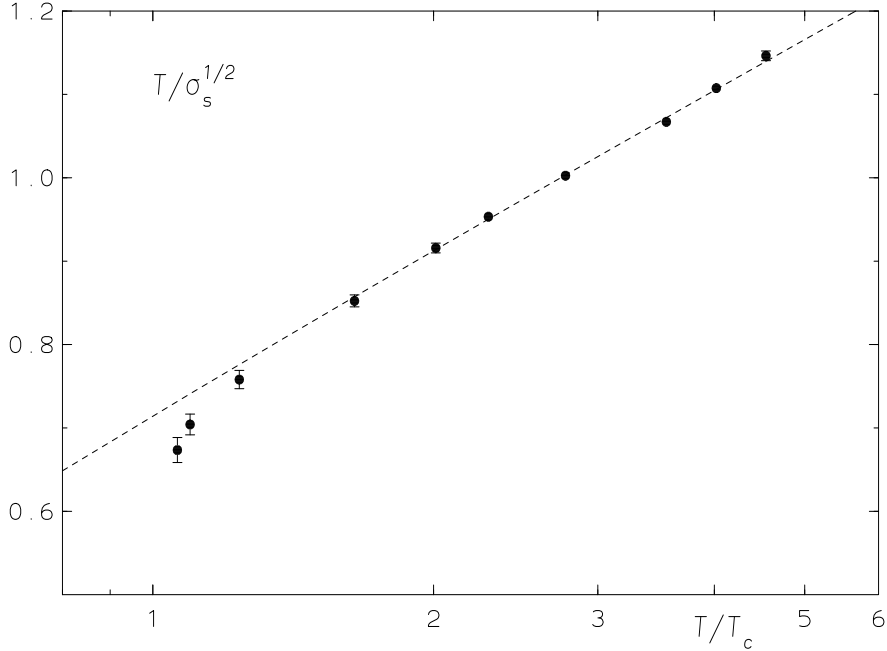,width=90mm,height=120mm,angle=-90}
\end{center}
\caption{The temperature over the square root of the spatial string tension
 versus $T/T_c$ for $SU(3)$. The dashed line shows the two-loop-fit 
according to Eq.~\protect\ref{sigmae}.}
\label{fig:string}
\end{figure}

We have calculated $\sigma_s$ at several values of the temperature on our
lattices with temporal extent $N_\tau = 8$. At each coupling we have
calculated smeared Wilson loops using the method of Ref. \cite{Balxx} 
up to size $(R,S)=(12,16)$, taking on-axis as well as off-axis 
separations for $R$. We have
analyzed these on 1500 configurations and extracted the static potentials
$V(R)$ from the logarithm of ratios of Wilson loops\footnote{Details on the
procedure to extract the heavy quark potential can be found in
Refs.~\cite{Balxx,sst3}.} at each value of the temperature. The spatial string
tension has been obtained from a
fit of these potentials with an ansatz including a linearly rising term
and a Coulomb term. The results for $T/\sqrt{\sigma_s(T)}$
are plotted in Fig.~\ref{fig:string}. They show that the spatial
string tension increases somewhat slower than linear and that the correction
is compatible with being logarithmic in $T/T_c$.
We therefore compare the temperature dependence with the expected behaviour
\beqn
\sqrt{\sigma_s(T)} = c g^2(T) T
\label{sigmae}
\eqn
and determine the parameter $c$ and $g^2(T)$ from a 2-parameter fit, where
we use for $g^2(T)$ the form given by the two-loop renormalization
group equation 
\beqn
g^{-2} (T) = 2b_0 \ln {T \over \Lambda_\sigma} + {b_1 \over b_0} \ln \biggl(
2 \ln{T \over \Lambda_\sigma}\biggr)~~,
\label{gtfit}
\eqn
with a free $\Lambda_\sigma$-parameter. This fit yields
\begin{eqnarray}
c &=& 0.566 \pm 0.013 ~~,
\nonumber \\
{\Lambda_\sigma \over T_c} &=& 0.104 \pm 0.009~~,
\label{sigmafit}
\end{eqnarray}
which is compatible with our earlier determination \cite{sst3}. We note that
these parameters correspond to a coupling $g^2(5T_c) = 1.5$ which is
consistent with the value estimated from the behaviour of the electric and
magnetic condensates at high temperature.

%%%%%%%%%%%%%%%%%%%%%%%%%%%%%%%%%%%%%%%%%%%%%%%%%%%%%%%%%%%%%%%%%%%%%%%%%%%%%%

\section{Conclusions}

%%%%%%%%%%%%%%%%%%%%%%%%%%%%%%%%%%%%%%%%%%%%%%%%%%%%%%%%%%%%%%%%%%%%%%%%%%%%%%
We have systematically analyzed the cut-off dependence of thermodynamic
observables of the $SU(3)$ gauge theory at finite temperature.  The high
statistics evaluation of the energy density and pressure on lattices with 
temporal extent $N_\tau= 4,$ 6 and 8 could be used to extrapolate these
quantities to the continuum limit. We find that energy density, entropy
density and pressure do approach the high temperature ideal gas limit from
below. At temperatures of about $5T_c$ deviations from the ideal gas limit
are still about 15\%. In fact, the approach to this limit is rather slow,
which is in agreement with the expectation that the functional dependence of
thermodynamic observables in this regime is controlled by a 
running coupling which varies with temperature only
logarithmically. The analysis of the bulk thermodynamics as well as other
observables like the spatial string tension in terms of leading order 
perturbative relations suggests that this running coupling is larger than 
unity even at $T\sim 5T_c$.

Within the Wilson formulation of $SU(N)$ gauge
theories thermodynamic observables show a cut-off dependence which is
$O((aT)^2)$ in the high temperature limit. This has been used in our
extrapolations to the continuum limit. The cut-off dependence can be 
reduced if one replaces the standard Wilson action by an improved version
which, for instance, could be of the form of a Symanzik improved action
\cite{Bei95,Sym83} or a perfect action \cite{Has94}. 
This will immediately lead to a reduction of the cut-off dependence of
thermodynamic observables in the high temperature ideal gas limit
($T\rightarrow \infty$)
\cite{Bei95}. In how far these actions will lead to a reduction of finite
cut-off effects also at finite temperatures, in particular for temperatures
close to $T_c$, remains to be seen.

\medskip
\noindent
{\large \bf Acknowledgements:}

\noindent
The work has been supported by the DFG under grant Pe 340/3-3.
It would not have been possible without
the 256-node QUADRICS parallel computer funded by the DFG under contract
no. Pe 340/6-1 for the DFG-Forschungsschwerpunkt "Dynamische Fermionen".
We thank M. Plagge for the smooth running of this facility.

%\vfill \eject

\newpage
%%%%%%%%%%%%%%%%%%%%%%%%%%%%%%%%%%%%%%%%%%%%%%%%%%%%%%%%%%%%%%%%%%%%%%%%%%%%%%

\section{Appendix}

%%%%%%%%%%%%%%%%%%%%%%%%%%%%%%%%%%%%%%%%%%%%%%%%%%%%%%%%%%%%%%%%%%%%%%%%%%%%%%

In the simulations we have used
a Cabibbo-Marinari pseudo-heatbath algorithm 
with Kennedy-Pendleton
updating in the $SU(2)$ subgroups. Every heatbath
step was supplemented by 4 to 9 overrelaxed updates, depending on
the $\beta$ value. This then constitutes one sweep. The accumulated
statistics on the asymmetric lattices amounts to 20 000 to 40 000 
sweeps after thermalization,
chosen according to the needed precision in the plaquette data.
On the symmetric lattices, between 5000 and 10000 sweeps were
sufficient.
Plaquettes and Polyakov loops were measured every sweep.
For the integrated autocorrelation length, we found a maximum of
250 sweeps for the Polyakov loop right at $T_c$, dropping
quickly to e.g. about 6 sweeps at $1.2 T_c$. Autocorrelation times
for the plaquette are considerably smaller. The errors on these
observables include the integrated autocorrelation lengths.
The Wilson loops were measured every 10th sweep. Errors on the
local potentials were determined by jackknife with a block length
of 100 measurements.

In the tables 
we give the plaquette expectation values $P_0$ calculated on
symmetric lattices of size $16^4$ and $32^4$ as well as the spatial
($P_\sigma$) and temporal ($P_\tau$) plaquette expectation values on
$16^3\times 4$ and $32^3\times 6$, 8 lattices. In actual calculations of
the pressure and energy density (Eqs.~\ref{freelat} and
\ref{deltalat}) we only need the sum ($P_\sigma + P_\tau$). As both
quantities are
strongly correlated, the error on the sum is the same as on the
individual quantities. The same is true for the error on the difference
($P_\sigma - P_\tau$).

%\vfill\eject
\begin{table}
%%%%%%%%%%%%%%%%%%%%%%%%%%%%%%%%%%%%%%%%%%%%%%%%%%%%%
%        16^3 * 4      und      16^4                %
%%%%%%%%%%%%%%%%%%%%%%%%%%%%%%%%%%%%%%%%%%%%%%%%%%%%%
\begin{center}
\begin{tabular}{|c||l|l||l|}
\hline
        & \multicolumn{2}{ c|| }{$16^3 \times 4$}
        & \multicolumn{1}{ c|  }{$16^4$}           \\
\hline
$\beta$  & \multicolumn{1}{ c|  }{$1 - P_{\sigma}$}
        & \multicolumn{1}{ c|| }{$1 - P_{\tau}$}
        & \multicolumn{1}{ c|  }{$1 - P_0$}       \\
\hline
5.6500  &  0.537598  ( 37)  &  0.537666  ( 34)  &  0.537642  (36) \\
5.6600  &  0.540079  ( 37)  &  0.540150  ( 37)  &  0.540073  (38) \\
5.6700  &  0.542559  ( 36)  &  0.542634  ( 36)  &  0.542401  (43) \\
5.6800  &  0.545150  ( 58)  &  0.545288  ( 63)  &  0.544748  (35) \\
5.6825  &  0.545780  ( 72)  &  0.545933  ( 82)  &  0.545345  (42) \\
5.6850  &  0.546630  (111)  &  0.546835  (130)  &  0.545862  (44) \\
5.6900  &  0.548928  (153)  &  0.549374  (180)  &  0.547021  (32) \\
5.6925  &  0.549954  (117)  &  0.550509  (139)  &  0.547564  (37) \\
5.6950  &  0.551210  (115)  &  0.551906  (138)  &  0.548088  (34) \\
5.7000  &  0.552969  ( 63)  &  0.553848  ( 76)  &  0.549214  (29) \\
5.7050  &  0.554302  ( 47)  &  0.555278  ( 52)  &  0.550205  (32) \\
5.7100  &  0.555517  ( 33)  &  0.556565  ( 37)  &  0.551265  (33) \\
5.7200  &  0.557530  ( 29)  &  0.558687  ( 31)  &  0.553357  (32) \\
5.7300  &  0.559380  ( 27)  &  0.560603  ( 28)  &  0.555353  (30) \\
5.7500  &  0.562804  ( 24)  &  0.564129  ( 25)  &  0.559121  (41) \\
5.7700  &  0.565953  ( 22)  &  0.567330  ( 22)  &  0.562692  (27) \\
5.8000  &  0.570263  ( 20)  &  0.571708  ( 20)  &  0.567651  (21) \\
5.8500  &  0.576908  ( 18)  &  0.578413  ( 18)  &  0.575115  (33) \\
5.9000  &  0.582987  ( 17)  &  0.584558  ( 17)  &  0.581792  (21) \\
5.9500  &  0.588681  ( 16)  &  0.590262  ( 16)  &  0.587977  (19) \\
6.0000  &  0.594072  ( 16)  &  0.595669  ( 16)  &  0.593678  (24) \\
6.1000  &  0.604114  ( 15)  &  0.605728  ( 14)  &  0.604115  (16) \\
6.2000  &  0.613385  ( 14)  &  0.615010  ( 14)  &  0.613647  (20) \\
6.3000  &  0.622028  ( 13)  &  0.623652  ( 13)  &  0.622450  (14) \\
6.3500  &  0.626162  ( 13)  &  0.627779  ( 13)  &  0.626610  (10) \\
6.4000  &  0.630191  ( 13)  &  0.631780  ( 13)  &  0.630663  (19) \\
6.5000  &  0.637846  ( 12)  &  0.639439  ( 12)  &  0.638411  (20) \\
6.6000  &  0.645093  ( 12)  &  0.646685  ( 12)  &  0.645712  (17) \\
6.7000  &  0.652020  ( 12)  &  0.653579  ( 11)  &  0.652635  (16) \\
6.8000  &  0.658610  ( 12)  &  0.660146  ( 11)  &  0.659236  ( 7) \\
\hline
\end{tabular}
\end{center}
\caption{The plaquette values on $16^3\times 4$ and $16^4$ lattices.}
\label{tab:plaq16}
\end{table}

%\newpage
\hoffset=-1.4truecm
\begin{table}
%\begin{center}
\begin{tabular}{|c||l|l||l|l||l|}
%%%%%%%%%%%%%%%%%%%%%%%%%%%%%%%%%%%%%%%%%%%%%%%%%%%%%
%     32^3 * 6 ,  32^3 * 8    und    32^4           %
%%%%%%%%%%%%%%%%%%%%%%%%%%%%%%%%%%%%%%%%%%%%%%%%%%%%%
%

\hline
        & \multicolumn{2}{ c|| }{$32^3 \times 6$}
        & \multicolumn{2}{ c|| }{$32^3 \times 8$}
        & \multicolumn{1}{ c|  }{$32^4$}           \\
\hline
$\beta$  & \multicolumn{1}{ c|  }{$1 - P_{\sigma}$}
        & \multicolumn{1}{ c|| }{$1 - P_{\tau}$}
        & \multicolumn{1}{ c|  }{$1 - P_{\sigma}$}
        & \multicolumn{1}{ c|| }{$1 - P_{\tau}$}
        & \multicolumn{1}{ c|  }{$1 - P_0$}   \\
\hline
5.800  &  0.5676489  ( 55)  &  0.5676526  ( 55)  &
&
             &  0.5676510  (205) \\
5.850  &  0.5751411  ( 58)  &  0.5751470  ( 58)  &
&
             &  0.5751226  ( 54) \\
5.870  &  0.5779126  ( 63)  &  0.5779266  ( 59)  &
&
             &  0.5778923  ( 54) \\
5.890  &  0.5806664  (255)  &  0.5807005  (340)  &
&
             &  0.5805461  ( 44) \\
5.895  &  0.5814918  (272)  &  0.5815575  (332)  &
&
             &  0.5811950  ( 46) \\
5.900  &  0.5822724  (193)  &  0.5823817  (233)  &
&
             &  0.5818383  ( 49) \\
5.910  &  0.5836330  ( 99)  &  0.5837652  (100)  &
&
             &  0.5831025  ( 46) \\
5.925  &  0.5855217  ( 42)  &  0.5856752  ( 43)  &
&
             &  0.5849659  ( 46) \\
5.950  &  0.5885074  ( 77)  &  0.5886900  ( 80)  &
&
             &  0.5879738  ( 40) \\
5.980  &  0.5919147  ( 34)  &  0.5921142  ( 34)  &
&
             &  0.5914373  ( 39) \\
6.000  &  0.5941102  ( 47)  &  0.5943184  ( 47)  &  0.5936825  (39)
&  0.59
36859  (39)  &  0.5936846  ( 39) \\
6.040  &                    &                    &  0.5980109  (34)
&  0.59
80153  (36)  &  0.5979958  ( 40) \\
6.060  &                    &                    &  0.6001360  (56)
&  0.60
01521  (73)  &  0.6000816  ( 34) \\
6.065  &                    &                    &  0.6006754  (60)
&  0.60
06988  (72)  &  0.6005991  ( 35) \\
6.070  &                    &                    &  0.6012168  (43)
&  0.60
12460  (46)  &  0.6011049  ( 36) \\
6.080  &                    &                    &  0.6022476  (36)
&  0.60
22829  (42)  &  0.6021223  ( 38) \\
6.100  &  0.6043655  ( 34)  &  0.6045954  ( 34)  &  0.6042724  (40)
&  0.60
43161  (40)  &  0.6041315  ( 38) \\
6.120  &                    &                    &  0.6062390  (38)
&  0.60
62881  (38)  &  0.6060953  ( 33) \\
6.140  &                    &                    &  0.6081606  (37)
&  0.60
82154  (37)  &  0.6080241  ( 35) \\
6.200  &  0.6137493  ( 42)  &  0.6139905  ( 40)  &  0.6137373  (35)
&  0.61
37980  (35)  &  0.6136303  ( 32) \\
6.300  & 0.6224639  ( 40)   &  0.6227104  ( 39)  &  0.6224809  (33)
&  0.62
25500  (33)  &  0.6224187  ( 30) \\
6.350  &  0.6266149  ( 35)  &  0.6268581  ( 35)  &
&
             &  0.6265895  ( 30) \\
6.400  &  0.6306247  ( 38)  &  0.6308739  ( 37)  &  0.6306677  (32)
&  0.63
07377  (32)  &  0.6306291  ( 28) \\
6.550  &  0.6420149  ( 42)  &  0.6422645  ( 41)  &  0.6420717  (30)
&  0.64
21439  (30)  &  0.6420618  ( 26) \\
6.650  &  0.6491184  ( 31)  &  0.6493655  ( 31)  &  0.6491777  (30)
&  0.64
92498  (29)  &  0.6491831  ( 19) \\
6.800  &  0.6591334  ( 47)  &  0.6593798  ( 46)  &  0.6592005  (28)
&  0.65
92707  (27)  &  0.6592132  ( 18) \\
7.000  &  0.6714649  ( 31)  &  0.6717057  ( 30)  &  0.6715327  (27)
&  0.67
16058  (27)  &  0.6715564  ( 16) \\
7.100  &                    &                    &  0.6773142  (27)
&  0.67
73845  (26)  &  0.6773391  ( 16) \\
7.200  &                    &                    &  0.6828632  (26)
&  0.68
29342  (25)  &  0.6828924  ( 16) \\
\hline
\end{tabular}
%\end{center}
\caption{The plaquette values on $32^3\times 6$ and 8 and $32^4$ lattices.}
\label{tab:plaq32}
\end{table}


\newpage
\begin{thebibliography}{99}
\bibitem{first} L.D. McLerran and B. Svetitsky, Phys. Lett. B98 (1981) 195; \\
J. Kuti, J. Polonyi and K. Szlachanyi, Phys. Lett. B98 (1981) 199;\\
J. Engels, F. Karsch, I. Montvay and H. Satz, Phys. Lett. B101 (1981) 89.
\bibitem{Linde} A.D. Linde, Phys. Lett. B96 (1980) 289.
\bibitem{Arnold} P. Arnold and C.-X. Zhai, Phys. Rev. D50 (1994) 7603.
\bibitem{Zhai} C.-X. Zhai and B. Kastening, Phys. Rev. D52 (1995) 7232.
\bibitem{Braaten} E. Braaten and R. Pisarski, Nucl. Phys. B 337 (1990) 569.
\bibitem{Eng95} J. Engels, F. Karsch and K. Redlich, Nucl. Phys. B 435 (1995)
295.
\bibitem{earlier} F. R. Brown et al., Phys. Rev. Lett. 61 (1988) 2058.
\bibitem{Wil74} K.G. Wilson, Phys. Rev. D 10 (1974) 2445.
\bibitem{Eng82} J.Engels, F.Karsch and H.Satz, Nucl. Phys. B 205 [FS5] (1982)
239.
\bibitem{Eng90} J. Engels, J. Fingberg, F. Karsch, D. Miller and M. Weber,
Phys. Lett. B 252 (1990) 625.
\bibitem{eosprl} G. Boyd, J. Engels, F. Karsch, E. Laermann, C. Legeland,
M. L\"utgemeier and B. Petersson, \PRL 75(1995) 4169.
\bibitem{Akemi} K. Akemi et al., Phys. Rev. Lett. 71 (1993) 3063.
\bibitem{Kuti} S.A. Gottlieb, J. Kuti, D. Toussaint, A.D. Kennedy, S.
Meyer, B.J. Pendleton and R.L. Sugar, Phys. Rev. Lett. 55 (1985) 1958.
\bibitem{Christ} N.H. Christ and A.E. Terrano, Phys. Rev. Lett. 56 (1986) 111.
\bibitem{Iwa92}  M. Fukugita, M. Okawa and A. Ukawa, Nucl. Phys. B337 (1990)
181; \\
Y. Iwasaki, K. Kanaya, T. Yoshi\'e, T. Hoshino, T.
Shirakawa, Y. Oyanagi, S. Ichii and T. Kawai, Phys. Rev. D 46 (1992) 4657.
\bibitem{Ferrenberg} M.~Falconi, E.~Marinari, M.L.~Paciello, 
G.~Parisi and B.~Taglienti, \PL 108B (1982) 331;
E.~Marinari, \NP B235 (1984) 123;\\
G.~Bhanot, S.~Black, P.~Carter and R.~Salvador, \PL 183B (1986) 331;\\
A.M. Ferrenberg and R.H. Swendsen,  Phys. Rev. Lett. 61
(1988) 2635; 63 (1989) 1195.
\bibitem{Balxx} G. Bali and K. Schilling, Phys. Rev. D46 (1992) 2636.
\bibitem{Fin93} J. Fingberg, U. Heller and F. Karsch, Nucl. Phys. B 392
(1993) 493.
\bibitem{Parisi} G. Parisi, Proceedings of the XX$^{th}$ Conference on High
Energy Physics, Madison 1980; \\
F. Karsch and R. Petronzio, Phys. Lett. 139B (1984) 403; \\
A.X. El-Khadra, G. Hockney, A.S. Kronfeld and P.B. Mackenzie, Phys. Rev.
Lett. 69 (1992) 729.
\bibitem{Bei95} B. Beinlich, F. Karsch and E. Laermann, {\it Improved
Actions for QCD Thermodynamics on the Lattice}, BI-TP 95/33, to be
published in Nucl. Phys. B.
\bibitem{Kar82} F. Karsch, Nucl. Phys. B205 [FS 5] (1982) 285.
\bibitem{Leu92} H. Leutwyler, in Proceedings of the Conference {\it QCD - 20
years later}, (Edts. P.M. Zerwas and H.A. Kastrup), World Scientific 1993,
p. 693.
\bibitem{Kap79} J. I. Kapusta, Nucl. Phys. B148 (1979) 461.
\bibitem{emcondens} V.K. Mitrjushkin, A.M. Zadorozhny and G.M. Zinovjev,
Phys. Lett. B215 (1988) 371; \\
S.H. Lee, Phys. Rev. D40 (1989) 2485; \\
H.G. Dosch, H.J. Pirner and Yu.A. Simonov, Phys. Lett. B349 (1995) 335. 
\bibitem{Hel85}  U. Heller and F. Karsch, Nucl. Phys. B251[FS13] (1985) 254.
\bibitem{sst2} L. K\"arkk\"ainen, P. Lacock,
D.E. Miller, B. Petersson and T. Reisz, Phys. Lett. B312 (1993) 173; \\
G.S. Bali, J. Fingberg, U.M. Heller, F. Karsch and K. Schilling,
Phys. Rev. Lett.  71 (1993) 3059.
\bibitem{sst3} F. Karsch, E. Laermann and M. L\"utgemeier, Phys. Lett.
B346 (1995) 94.
\bibitem{Sym83} K. Symanzik, Nucl. Phys. B226 (1983) 187 and Nucl. Phys.
B226 (1983) 205.
\bibitem{Has94} P. Hasenfratz and F. Niedermayer, Nucl. Phys. B414 (1994)
785; \\
T. DeGrand, A. Hasenfratz, P. Hasenfratz and F. Niedermayer,
Nucl. Phys. B454 (1995) 587 and 615.
\end{thebibliography}
\end{document}